# Study of SiO$_2$-PbO-CdO-Ga$_2$O$_3$ glass system for mid-infrared optical elements

Xavier Forestier[a, b], Jarosław Cimek[a, c], Ireneusz Kujawa[a], Rafał Kasztelanic[a, c], Dariusz Pysz[a], Krzysztof Orliński[a], Ryszard Stępień [a], Ryszard Buczyński[a,c]

[a] Institute of Electronic Materials Technology, Wolczynska 133, 01-919 Warsaw, Poland
[b] Warsaw University of Technology, Faculty of Physics, Koszykowa 75, 00-662 Warsaw, Poland
[c] University of Warsaw, Faculty of Physics, Pasteura 5, 02-093 Warsaw, Poland

**Abstract:**

Glasses based on SiO$_2$-PbO-CdO-Ga$_2$O$_3$ system have been studied for the first time for fabrication of mid-infrared optical elements. Gallium oxide concentration was gradually increased, replacing silicon dioxide, for different cadmium and lead oxide content. The thermal and optical properties were investigated for different compositions. It was observed that the thermal stability, refractive index, and the transmission in the infrared range increased with increase of gallium and lead concentrations. The most thermally stable glass composition was selected for fabrication of optical elements such as optical fibers. We also successfully fabricated mid-infrared lenses by hot embossing for potential application in compact gas detectors.

**Highlights:**

- Thermal and optical properties of SiO$_2$-PbO-CdO-Ga$_2$O$_3$ glass system were investigated.
- Substitution of silicon dioxide with gallium oxide decreases glass susceptibility to crystallization without significant change in refractive index.
- The most thermally stable glass composition of 28SiO$_2$-45PbO-15CdO-12Ga$_2$O$_3$ is suitable for the fabrication of lenses by the hot embossing method and for the fabrication of optical fibers.

### 1. Introduction:

Recently, development of optical components dedicated for mid-infrared wavelengths has attracted considerable interest due to market demands. Indeed, mid-infrared wavelength range, generally defined between 2 µm to 15 µm, finds many applications in environmental monitoring, food safety, biomedical sensing or in the defence industry. Polymer materials cannot be used in this range due to very high attenuation. Glasses for infrared optics like optical fibers or lenses require the use of materials with appropriate composition to ensure high transmission in this range of wavelengths, as well as good mechanical and physicochemical properties including good chemical durability. Moreover the thermal processes involved in fiber drawing using the stack and draw method [1] and in lenses molding by the hot embossing method [2-3] require the glass to be stable regarding crystallization over a wide range of temperatures. The optical and thermo-physical properties of the main type of glasses used in infrared optics are presented in Table 1.

Table 1 Main properties of the families of glass used in infrared optic

| | | n at 1.55μm | $n_2 \times 10^{20}$ (m²/W) | Transmission (μm) | CTE $\alpha_{20/300}$ ($10^{-7}$·K$^{-1}$) | $T_g$ (°C) | Ref |
|---|---|---|---|---|---|---|---|
| Silicates | silica | 1.44 | 2.73 | 0.2 to 2.7 | 5.9 | ~1200 | [4] |
| | borosilicate (BK7) | 1.50 | 3.60 | 0.3 to 2.5 | 73 | ~498 | [5] |
| | lead silicate (SF57) | 1.80 | 41 | 0.41 to 4.5 | 92 | ~414 | |
| | HMOG with silica | 1.8-2.2 | 30-100 | 0.4 to 5.5 | 70 to 100 | 400 to 500 | [6] [7] |
| Fluorides | ZBLAN | 1.51 | 3.3 | 0.3 to 8 | 200 | ~265 | [8] |
| Heavy metal oxides glass | | 1.8-2.2 | 30-100 | 0.4 to 7 | 80 to 120 | 230 to 400 | [9] |
| Chalcogenides | $As_2S_3$ | 2.45 | 400-600 | 0.62 to 11.53 | 230 | ~185 | [10] |
| | $As_2Se_3$ | 2.81 | 1400-3000 | 0.85 to 17.5 | 250 | ~178 | |
| | $Te_{20}As_{40}Se_{40}$ | 2.9 | >2000 | 1.23 to 18.52 | 230 | ~140 | |

Silica glasses are widely used in optics for telecommunication applications considering their excellent transmission and their low attenuation losses in the visible and the near infrared. Silica-based fibers can also be doped with transition metals or rare earth ions for optical amplifier application, doped with erbium or neodymium for example [11-13]. However, their increasing absorption above 2 μm excludes applications in the mid-infrared.

Soft glasses and particularly heavy metals oxides, chalcogenides and fluoride glasses present the advantage to have an extended transmission in the infrared and have higher linear and nonlinear refractive indices, which makes them particularly interesting for the development of optical components in the infrared. They also present the advantage to have lower processing temperatures.

Fluoride glasses, with their extremely broad transmission from ultraviolet to the mid-infrared, up to 8 μm, have a linear and nonlinear refractive indices similar to silicates. Moreover, their high rare-earth solubility and their low attenuation allow fiber lasers application [14-15] or generation of supercontinuum with ultrashort high power pulses [16]. On the other hand, this family of glasses has poor chemical durability if compared to heavy metal oxides ones.

Chalcogenide glasses offer a large transmission window in the infrared, high linear and nonlinear refractive indices as well as low phonon energy, which makes them excellent host materials for amplifier and laser applications [17], despite their rather low rare earth solubility [18]. Their unique properties of high transmission window also makes them appropriate for application in IR optics such as thermal imaging [19], metrology [20], military defense systems or medicine [21]. Sulfur and selenium based glasses have a transparency up to 10 μm and 16 μm respectively [22]. Tellurium based glasses, on the other hand, can exhibit a multiphonon cut-off wavelength up to over 24 μm due to Te higher atomic weight [23]. However, due to the strong metallic character of tellurium, they have a tendency to crystallize during thermal processing.

On the other hand, heavy metal oxides glasses (HMOG) have a limited transmission in the mid-IR up to 6-7 μm and a refractive index ranging from 1.8 to over 2.2 but have the advantage to have a good transparency starting in the visible and overall better mechanical properties [24]. They offer a good compromise between optical properties, cost and are relatively easy to synthesize. Lead gallium glass system, lacking typical glass formers, exhibits a low thermal stability. Previous studies have shown that the introduction of $SiO_2$ in lead-gallium glasses increases the crystallization resistance, due to the stabilizing role of $[SiO_4]$ tetrahedra in the glass network, while keeping the transmission over 4.5 μm [25]. In addition, gallium oxide dramatically improves the thermal stability of heavy metal oxides glasses against devitrification [26]. Incorporating $Ga_2O_3$ into glass compositions additionally increases the linear and nonlinear index of refraction [27-30]. Lead oxide is known to increase the refractive index due to its high optical density and large molecular weight. It also decreases the viscosity and the characteristic temperature due to the creation of non-bridging oxygen in silicate glasses, below 30% content. Above 30%, PbO starts to act as a glass former and enter the glass network with PbOx polyhedron, further decreasing the transition temperature [31-32]. Such multicomponent system allows to easily tune the thermomechanical and optical properties by modifying the concentration of the components [33].

The goal of the present work is to develop glasses with the appropriate rheological properties for the fabrication of optical elements such as micro-lenses and fiber optics with high transmittance in mid-infrared. We have chosen to study the glass system containing $SiO_2$-PbO-CdO-$Ga_2O_3$ (named SPCG) that was not considered before. In this aim, we studied in the proposed $SiO_2$-PbO-CdO-$Ga_2O_3$ system, the impact of the substitution of silica by gallium oxide on the thermal and optical properties for different lead oxide and cadmium oxide contents. We consider the most stable glasses against crystallization for the fabrication of optical fibers by the stack and draw method as well as lenses by the hot embossing method.

## 2. Synthesis of $SiO_2$-PbO-CdO-$Ga_2O_3$ glass composition

The glasses were prepared by the conventional melt-quenching technique using analytical grade substrates $SiO_2$, CdO, $Ga_2O_3$, PbO and $As_2O_3$ as fining agent [32]. Fining agents are used to obtain glasses with high optical quality. $As_2O_3$ is used in combination with oxidant, lead nitrates in this case, to convert $As_2O_3$ to $As_2O_5$. During the fining, at higher temperature, pentavalent arsenic will release $O_2$ in the melt, promoting gas removal from the molten glass, increasing the homogeneity. The substrates were weighed and ground to obtain homogenous powder followed by drying at 150°C for twelve hours. Batches were finally melted in alumina crucibles for 2 hours, at a temperature ranging from 1000°C to 1100°C, depending on the composition, in a resistance furnace. Melts have been mechanically stirred several times to obtain a good homogeneity and have then been poured into a preheated graphite mold before being annealed, from 460 degrees to the room temperature, at a rate of 0.5°C/min. In order to investigate the influence of the composition on the properties and to develop thermally stable glasses for mid-infrared optical elements, 3 series of glasses have been synthesized: series D, with the general composition (40-x)$SiO_2$-35PbO-25CdO-x$Ga_2O_3$, series B, (40-x)$SiO_2$-40PbO-20CdO-x$Ga_2O_3$ and series C, (40-x)$SiO_2$-45PbO-15CdO-x$Ga_2O_3$, with x=0, 3, 6, 9, 12 and 15 %mol. The chemical compositions are presented in Table 2.

Table 2: Chemical composition (in mol%) of SPCG B, C and D series of glasses

| Oxide | Composition of SPCG glasses (mol%) | | | | | | | | | | | | | | | | | |
|---|---|---|---|---|---|---|---|---|---|---|---|---|---|---|---|---|---|---|
| | (40-x)$SiO_2$-40PbO-20CdO-x$Ga_2O_3$ | | | | | | (40-x)$SiO_2$-45PbO-15CdO-x$Ga_2O_3$ | | | | | | (40-x)$SiO_2$-35PbO-25CdO-x$Ga_2O_3$ | | | | | |
| | 0B | 1B | 2B | 3B | 4B | 5B | 0C | 1C | 2C | 3C | 4C | 5C | 0D | 1D | 2D | 3D | 4D | 5D |
| $SiO_2$ | 40 | 37 | 34 | 31 | 28 | 25 | 40 | 37 | 34 | 31 | 28 | 25 | 40 | 37 | 34 | 31 | 28 | 25 |
| PbO | 40 | 40 | 40 | 40 | 40 | 40 | 45 | 45 | 45 | 45 | 45 | 45 | 35 | 35 | 35 | 35 | 35 | 35 |
| CdO | 20 | 20 | 20 | 20 | 20 | 20 | 15 | 15 | 15 | 15 | 15 | 15 | 25 | 25 | 25 | 25 | 25 | 25 |
| $Ga_2O_3$ | - | 3 | 6 | 9 | 12 | 15 | - | 3 | 6 | 9 | 12 | 15 | - | 3 | 6 | 9 | 12 | 15 |

## 3. Measurement procedure of $SiO_2$-PbO-CdO-$Ga_2O_3$ glass

The viscosities were determined by a combination of hot stage microscopy (HSM), dilatometry ($\eta = 10^{11}$ P) and differential scanning calorimetry ($\eta = 10^{13.4}$ P). HSM consists of a cubic sample (dimension 4 x 4 x 4 mm) heated in a furnace with a 10°C/min heating rate and observed with an objective. The viscosities have been determined depending on the evolution of the shape of the samples. When the corners of the cube round out, the viscosity $\eta = 10^9$ P. As the temperature increases, the sample goes from a spherical shape at $T_{sph}$ ($\eta = 10^6$ P) to a hemispherical shape at $T_{hs}$ ($\eta = 10^4$ P) to finally spread at lower viscosity ($\eta = 10^2$ P) [34-35]. Characteristic temperatures ($T_c$, $T_{hs}$, $T_{sph}$, $T_{spr}$) were determined by the observation of the sample shape through the objective. Several measurements were performed and show an accuracy of ±5°C.

The linear coefficients of thermal expansions along with the dilatometric softening temperatures ($\eta = 10^{11}$ P), $T_{dsp}$, have been determined with a Bahr Thermoanalyse GmbH Dil801 dilatometer with 5°C/min heating rate, under air. The accuracy of the linear expansion coefficient is ±0.03.$10^{-6}$ K$^{-1}$, whereas the accuracy on the temperature measurement is ±1°C. Measurements were performed on rods with the dimension 4 x 4 x 30 mm.

Glass transition temperatures $T_g$, as well as the crystallization onset temperature $T_x$, were determined by differential scanning calorimetry (DSC) using STA 449 F1 Jupiter (NETZSCH) equipped with a platinum furnace. Samples of 60 mg were heated in alumina crucibles with 10 K/min heating rate under the flow of a mixture comprising argon (72 ml/min) and oxygen (18 ml/min), in the range 30-800°C. The thermal stability of the glasses can be evaluated by the difference $\Delta T = T_x - T_g$ introduced by Dietzel [36]. A glass is stable enough for fiber drawing if ΔT is above 100°C and if it has a good crystallisation resistance over a wide range of temperature. Indeed the various heat treatments involved in the production of optical elements can lead to the devitrification of the glass.

Additional crystallization tests have been carried out with an isothermal heat treatment, 20°C higher than the sphere creation temperature $T_{sph}$ (log η~4-5 P), for 2 hours and then cooled down to room temperature. Samples were then observed under microscope with a polarizing attachment to detect eventual crystals growth due to the heat treatment.

The density of the glasses has been determined at room temperature by the conventional Archimedes method, using distilled water as immersion liquid, at room temperature. From the density and the average molar weight of the glass, the molar volume has been calculated using the following formula:

$$V_m = \sum_{i=1}^{n} x_i . M_i / \rho \quad (cm^3 . mol^{-1}) \tag{1}$$

Where $x_i$ is the molar fraction and $M_i$ is the molecular weight of the $i^{th}$ component and ρ the glass density in g.cm$^{-3}$.

Glass transmittance measurements have been conducted with a spectrophotometer Bruker IFS 113V from 2 μm to 6 μm with 2 mm thickness polished samples.

The group refractive index has been measured using the Michelson interferometry technique with a Thorlabs CCS220 spectrometer from 200 nm to 1000 nm and an Avantes AvaSpec-NIR256-1.7 spectrometer from 1 μm to 1.7 μm. The systematic error calculated for the group refractive index measurement is ± 4.10$^{-4}$. The phase refractive indices (2) were determined from the Sellmeier coefficients, retrieved by fitting the first derivative of the Sellmeier equation (3) to the measured group refractive indices [37].

$$n(\lambda) = \sqrt{1 + \frac{B_1 \lambda^2}{\lambda^2 - C_1} + \frac{B_2 \lambda^2}{\lambda^2 - C_2} + \frac{B_3 \lambda^2}{\lambda^2 - C_3}} \tag{2}$$

Where n is the phase refractive index at wavelength λ in µm, $B_1, B_2, B_3$ and $C_1, C_2, C_3$ are constants determined by the fitting process.

$$N(\lambda) = n(\lambda) - \lambda \cdot \frac{dn(\lambda)}{d\lambda} = n(\lambda) + \frac{\lambda^2}{n(\lambda)} \cdot \sum_{i=1}^{n} \frac{B_i \cdot C_i}{(\lambda^2 - C_i)^2} \quad (3)$$

Where $N(\lambda)$ is the group refractive index, $n(\lambda)$ the phase refractive index, λ the wavelength in µm, Bi and Ci are constants determined by the fitting process.

The optical dispersion of the glasses have been evaluated from their respective Abbe V-number calculated from:

$$v_D = \frac{n_D - 1}{n_F - n_C} \quad (5)$$

With $n_D$, $n_F$, and $n_C$ respectively the indices of refraction for the sodium D line (589.3 nm), the hydrogen F line (486.1 nm) and the hydrogen C line (656.3 nm).

The material dispersion of the bulk glasses has then been calculated from the second derivative of the Sellmeier equation such as:

$$D(\lambda) = -\frac{\lambda}{c} \cdot \frac{dn^2}{d^2\lambda} \quad (6)$$

## 4. Results and discussion
### 4.1 DSC measurements

The DSC measurement results for the 3 series of glass are presented Figure 1. The transformation temperature $T_g$, crystallization onset temperature $T_x$ and the difference $\Delta T = T_x - T_g$ are summarized in Table 3.

Table 3 Characteristic temperatures of SPCG glasses determined by DSC

| Serie B | $T_g$ (°C) | $T_x$ (°C) | $\Delta T = T_x - T_g$ | Serie C | $T_g$ (°C) | $T_x$ (°C) | $\Delta T = T_x - T_g$ | Serie D | $T_g$ (°C) | $T_x$ (°C) | $\Delta T = T_x - T_g$ |
|---|---|---|---|---|---|---|---|---|---|---|---|
| SPCG0B | 455.3 | 613.5 | 158.2 | SPCG0C | 439.5 | 575 | 135.5 | SPCG0D | 459.5 | 581.7 | 122.2 |
| SPCG1B | 454.0 | 691.4 | 237.4 | SPCG1C | 443.5 | 596 | 152.5 | SPCG1D | 467.8 | 613.2 | 145.4 |
| SPCG2B | 460.3 | 660.9 | 200.6 | SPCG2C | 451.7 | 649.7 | 198.0 | SPCG2D | 474.0 | 617.9 | 143.9 |
| SPCG3B | 465.6 | 670.7 | 205.1 | SPCG3C | 460 | 652.7 | 192.7 | SPCG3D | 478.8 | 619 | 140.2 |
| SPCG4B | 468.1 | 670.5 | 202.4 | SPCG4C | 460.8 | 654.2 | 193.4 | SPCG4D | 480.3 | 677.5 | 197.2 |
| SPCG5B | 468.0 | 655.3 | 187.3 | SPCG5C | 464.5 | 652 | 187.5 | SPCG5D | 480.8 | 679.2 | 198.4 |

*Accuracy of the measurement was evaluated comparing the melting point of different pure metals in the range of measurement (100 - 900°C). The accuracy is ±0.3°C due to varying heat transfer conditions such as the position of the crucible in the heating chamber, the mass of the sample and the heating rate.

Glasses with 0 % and 3 % gallium in all series present multiple sharp crystallization peaks, characteristic of a high tendency to crystallize and point to the presence of multiple crystalline phases. The most stable glasses with 6 % to 15 % $Ga_2O_3$ are characterized by a very low intensity crystallization peak and a good nonisothermal crystallization resistance with ΔT ranging from 150 K to 205 K. Additional endothermal reactions of unknown nature are observed in the ΔT region. The thermal stability of all series increases with the substitution of $SiO_2$ with $Ga_2O_3$, as well as the glass transition temperature (Fig 2), ranging from 439°C to 480°C. We observe also that replacing cadmium oxide with lead oxide leads to a diminution of the characteristics temperatures. Indeed, an increase of PbO concentration instead of CdO, decreases the glass transition temperature.

The additional isothermal crystallization test confirmed that SPCG 3, 4 and 5 of each series are the most suitable to make optical elements and have been selected for further tests for hot embossing and fiber drawing applications. In these cases, no crystallites were observed under the polarizing microscope after 2 hours of heat treatment at $T_{sph}$ +20 K, in contrast with SPCG 0, 1 and 2 of each series, for which complete or partial crystallization was observed. We conclude that increasing $Ga_2O_3$ content with simultaneous reduction of $SiO_2$ leads to a large augmentation of the thermal stability of the glass, as shown in Table 3.

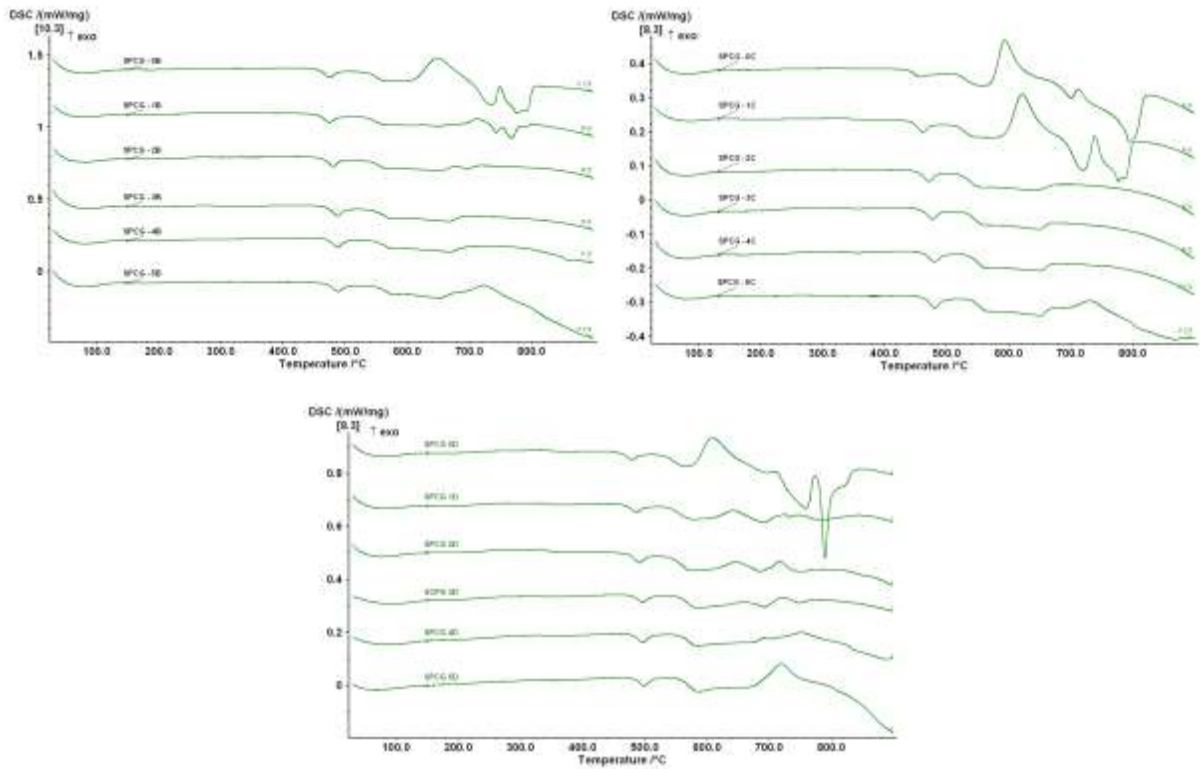

Figure 1 DSC curves for SPCG B series (left), SPCG C series (right) glasses and SPCG D series (down).

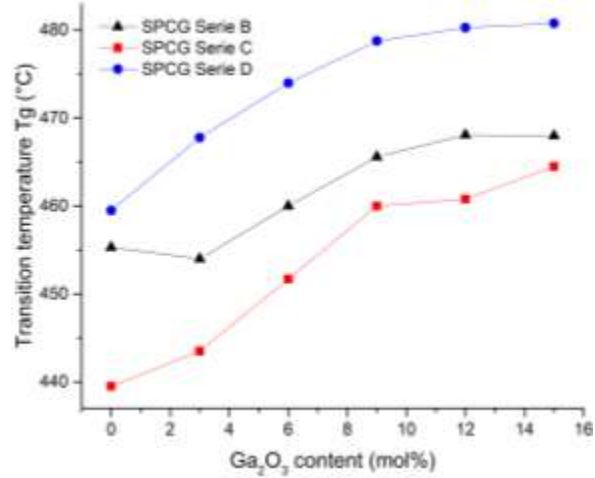

*Figure 2 Evolution of the glass transition temperature with the gallium oxide content.*

### 4.2 Viscosity measurements

The evolution of the temperature dependence of the viscosity is presented in Figure 3. The glasses containing 0 and 3% $Ga_2O_3$ concentration have shown a high tendency to devitrify, therefore, the viscosities at high temperature could not be obtained by hot stage microscopy due to the crystallization of the sample between the curvature and the sphere temperatures. Further increase of $Ga_2O_3$ concentration leads to a monotonic augmentation of the viscosity. Substituting PbO with CdO leads to an increase of the viscosity as well. Series D (35%PbO/25%CdO) presents the highest viscosities among all series of glass whereas series C (45%PbO/15%CdO) has the lowest. The viscosity increases with an increasing $Ga_2O_3$ and CdO content, which is consistent with previous studies [38] showing that an increasing lead oxide molar percentage decreases the viscosity of lead silicate glasses. The measured viscosity characteristics of glasses shows that they are technologically short, which makes fabrication of fibers by the stack and draw technique delicate. If we consider the glass processing viscosity between $\eta=10^4$ and $10^8$ Poises for the production of optical elements, the working temperature range is comprised between 525 ºC and 615 ºC (Fig.3). Due to the different nature of the measurements involved (as explained section 3), the viscosity curves, logarithmically fitted with the Levenberg-Marquardt algorithm, is added as a guide for the eyes to get the general trend of the different glass compositions.

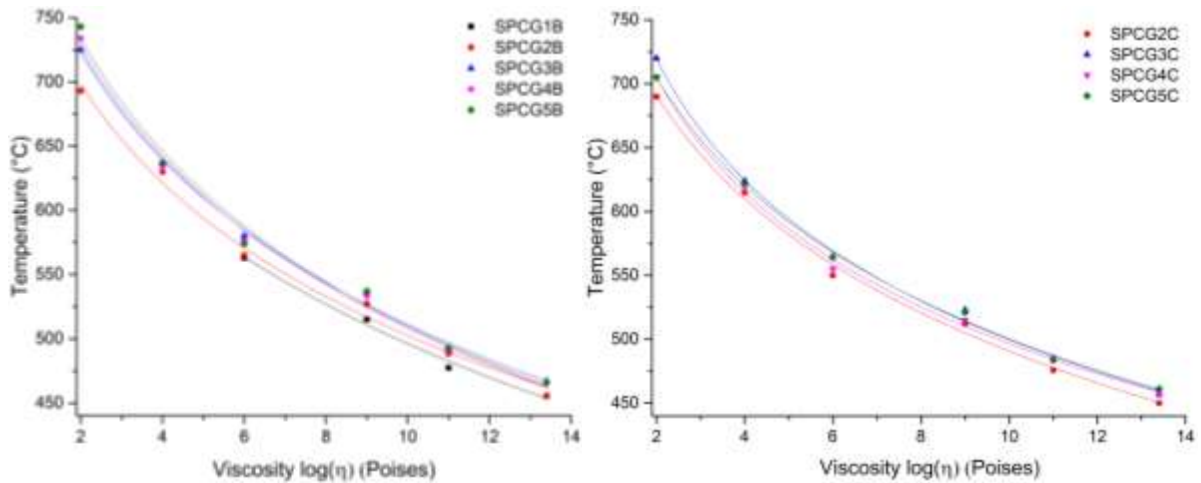

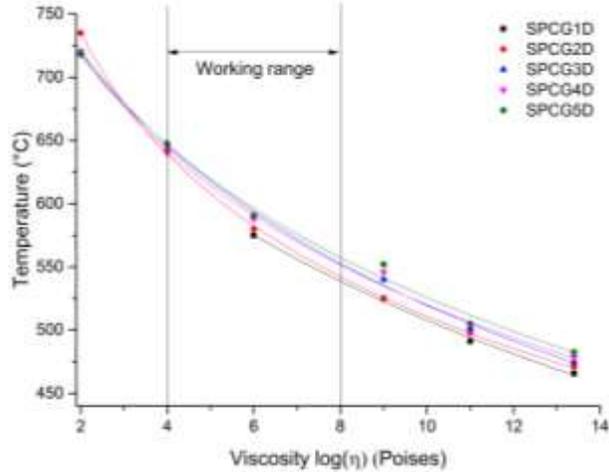

*Figure 3 Viscosity-Temperature relationship of SPCG glasses for B, C and D series*

### 4.3 Density and molar volume

The density increases with the increase of gallium oxide content (Fig.4). This is related to the substitution of lighter atoms of Si with heavier atoms of Ga in the network which leads to increase the average molecular weight of the glass. The density anomaly observed at 9% content of $Ga_2O_3$ disappears, if we calculate the molar volume. Indeed, the molar volume is more sensitive to structural changes than is density because it is normalized for atomic weights of different glass components. We observe an augmentation of the molar volume with an increasing $Ga_2O_3$ molar percentage (Fig. 5). In general, density and molar volume vary in opposite way, this atypical behavior between density and molar volume has been previously reported also for other glass systems [39-40]. In the present case, this may be attributed to the molecular weight of the glass increasing more rapidly than the density. The density of series B is much higher than the 2 other series leading to the lowest molar volume among the 3 series, which may point to a structural change at 40% PbO.

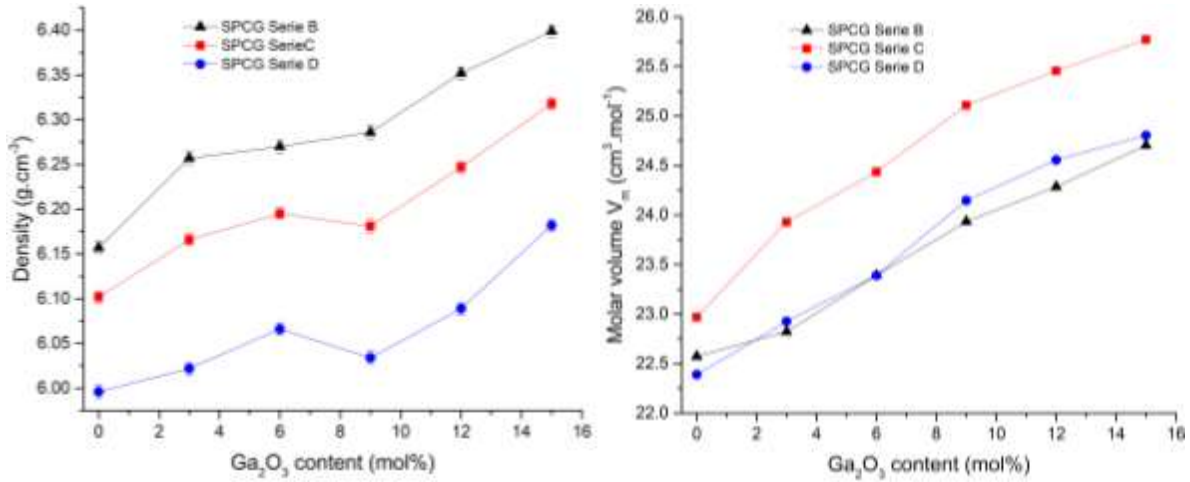

*Figure 4 Evolution of the density (left) and molar volume (right) of SPCG glasses with $Ga_2O_3$ content.*

Table 4 Properties of SPCG glasses series

| Glass | CTE $\alpha_{300}^{20}$ (10$^{-6}$K$^{-1}$) ±0.03 | $T_{dsp}$ (°C) ±1 | Characteristics temperatures in Hot Stage Microscopy (°C) ±5°C | | | | Cristallization | Density ρ (g/cm$^3$) ±0.007 | Molar Volume (cm$^3$/mol) ±7.10$^{-4}$ | Abbe Number | Refractive index $n_d$ |
|---|---|---|---|---|---|---|---|---|---|---|---|
| | | | Curvature $T_c$ | Sphere $T_s$ | Hemisphere $T_{hs}$ | Spreading $T_{spr}$ | | | | | |
| SPCG0B | 9.15 | 477.8 | 510 | Xsed | Xsed | Xsed | yes | 6.157 | 22.573 | - | - |
| SPCG1B | 8.95 | 477.3 | 515 | 563 | Xsed | Xsed | yes | 6.257 | 22.822 | 20.0 | 1.945 |
| SPCG2B | 8.53 | 488.4 | 527 | 565 | 630 | 693 | yes | 6.27 | 23.389 | 18.5 | 1.957 |
| SPCG3B | 8.26 | 491.4 | 537 | 580 | 637 | 725 | no | 6.286 | 23.934 | 19.7 | 1.975 |
| SPCG4B | 8.15 | 491 | 533 | 576 | 634 | 734 | no | 6.352 | 24.285 | 19.4 | 1.982 |
| SPCG5B | 8.2 | 493 | 537 | 574 | 637 | 743 | no | 6.399 | 24.705 | 18.6 | 1.991 |
| SPCG0C | 9.06 | 462.4 | Xsed | Xsed | Xsed | Xsed | yes | 6.141 | 22.969 | 20.5 | 1.965 |
| SPCG1C | 8.73 | 466.2 | 500 | Xsed | Xsed | Xsed | yes | 6.166 | 23.929 | 19.6 | 1.970 |
| SPCG2C | 8.59 | 475.8 | 512 | 550 | 615 | 690 | yes | 6.195 | 24.435 | - | - |
| SPCG3C | 8.4 | 484.8 | 523 | 565 | 624 | 720 | yes | 6.181 | 25.108 | 19.1 | 1.980 |
| SPCG4C | 8.26 | 484.4 | 515 | 555 | 620 | 705 | no | 6.247 | 25.455 | 18.6 | 1.992 |
| SPCG5C | 8.42 | 483.6 | 521 | 564 | 622 | 705 | no | 6.318 | 25.771 | 17.8 | 1.998 |
| SPCG0D | 9.27 | 483.6 | 545 | Xsed | Xsed | Xsed | yes | 5.996 | 22.389 | 21.4 | 1.940 |
| SPCG1D | 8.54 | 491.2 | 525 | 575 | Xsed | Xsed | yes | 6.022 | 22.926 | 21.2 | 1.944 |
| SPCG2D | 8.26 | 497.8 | 530 | 581 | 642 | 730 | yes | 6.066 | 23.391 | 20.4 | 1.948 |
| SPCG3D | 8.18 | 501.7 | 540 | 590 | 648 | 720 | no | 6.034 | 24.147 | 20.2 | 1.956 |
| SPCG4D | 8.51 | 498.3 | 546 | 588 | 640 | 720 | no | 6.089 | 24.556 | 19.5 | 1.970 |
| SPCG5D | 8.37 | 505.1 | 552 | 591 | 647 | 718 | no | 6.182 | 24.804 | 19.3 | 1.980 |

### 4.4 Optical Properties

The transmission measurements are presented in Fig. 5. All series of developed glasses show a broad spectral window in the mid-infrared, up to 5.0 µm. Increasing the gallium oxide content leads to shifting the multiphonon edge towards longer wavelengths, whereas the substitution of lead oxide by cadmium oxide lets the transmission in infrared almost unchanged. The intensity of the characteristic absorption peak related to the presence of hydroxyl ions OH$^-$, as well as the transmission cut-off wavelength, increase with the gallium content. We also noticed that the absorption peak intensity around 3 µm decreases with the substitution of PbO with CdO. This shows that lead oxide tends to retain more water than cadmium oxide. The strong water absorption is mostly due to the synthesis without protective atmosphere as well as the water contained in the initial components.

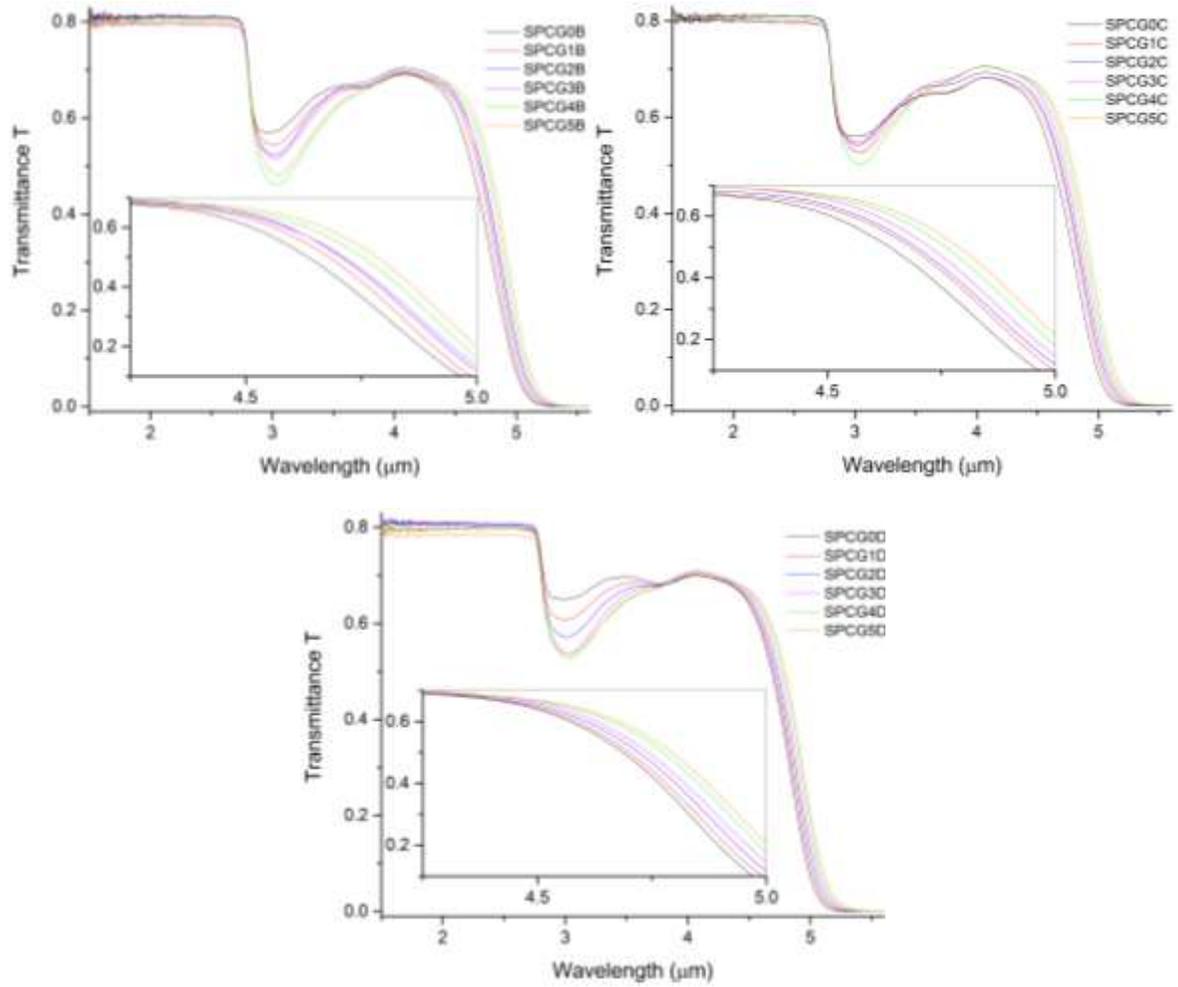

*Figure 5 Transmittance of SPCG glasses in the range 2 to 6 µm.*

The refractive indices have been calculated from the group refractive index, measured with a Michaelson interferometer as discussed section 3, from 500 to 1700 nm. The sodium D line (589.3 nm) has been used to compare the glasses and are ranging from 1.940 to 1.998 among the 3 series. The substitution of silica by gallium, as well as an increasing lead oxide content, lead to an augmentation of the refractive index. As shown in Figure 6, the highest refractive index is obtained for Series C with 45 %mol PbO ($n_D$=1.998) and the lowest for series D with 35 %mol PbO ($n_D$=1.940). The material dispersion D, calculated from the second derivative of the Sellmeier equation (Eq. 6), follows the exact same trend (Fig. 7). The evolution of the reciprocal relative dispersion, or Abbe V-number, with the composition is summarized in Table 4. The optical dispersion is high for all glasses and increases with the gallium percentage. The Sellmeier coefficients of the glasses are summarized in the annex 1.

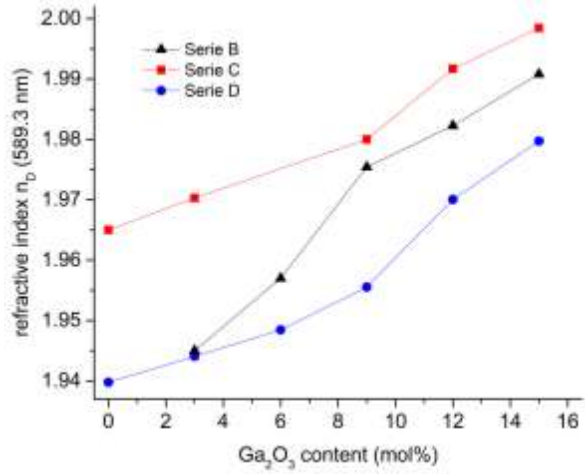

Figure 6 Evolution of the refractive index at 589.3 nm as a function of $Ga_2O_3$ content.

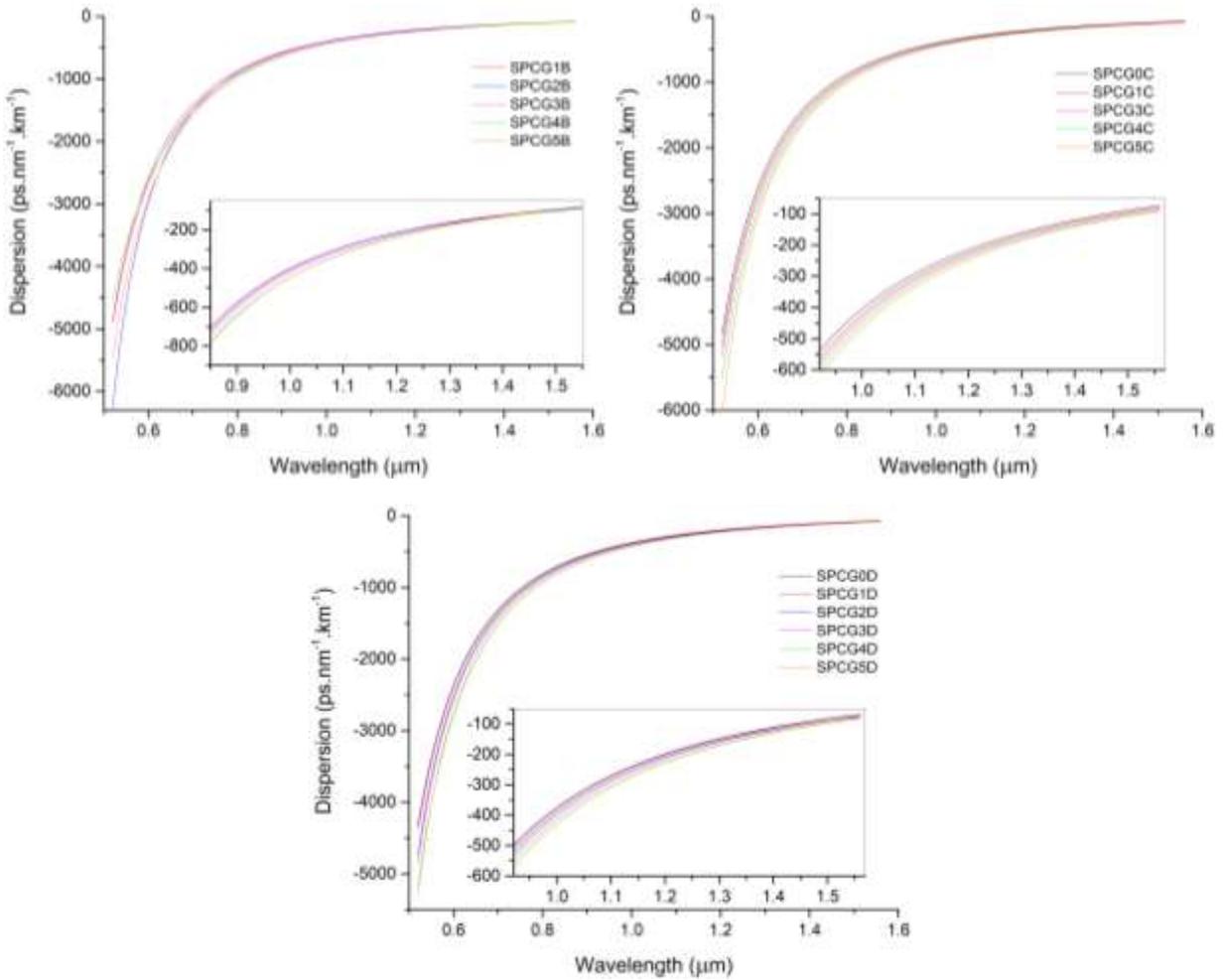

Figure 7 Dispersion parameter D as function of the wavelength for the 3 series of glasses.

## 5. Verification of use of selected PbO-SiO$_2$-CdO-Ga$_2$O$_3$ glasses for fiber and micro-optic development

Hot embossing process presents the advantage to be one of the lowest-cost methods for the fabrication of optical elements. It consists in an upper and lower stamps heated to the desired temperature [41]. When the optimal temperature of the glass is attained, the desired shape is finally transferred by pressure of the molding stamps. One of the most important steps to make lenses by the hot embossing technique is the choice of the mold material. Indeed, the coefficient of thermal expansion (CTE) of the glass must match the chosen mold to avoid cracking during the process. In addition, the glass must not adhere to the mold and has to be crystallization resistant. We have selected brass and cemented carbide for their very close CTE. These materials were successfully used as molds for hot embossing with heavy metal oxide glasses previously [42]. We've selected the glasses with the largest ΔT criterion (SPCG 3, 4, 5 of each series) which resisted to the isothermal crystallisation test (according to section 4.1) and have conducted preliminary embossing tests. SPCG4C glass was synthetized at increased volume without any changes of the optical and thermal properties which proves a good repeatability of the experiments. Different molding parameters have been tested until the obtention of a lens. Such parameter, beside the choice of the mold, are the molding temperature, the force applied and the molding duration. Based on the quality of the lens and the eventual presence of cracks or irregularities of the surface of the different lenses developed from the 3 series, we have finally chosen SPCG4C glass for its good thermal stability and its capacity to form good quality lenses. Indeed, other glasses could not be molded properly or showed cracks. The optimal parameters used for this composition were a temperature of 602ºC, the pressure force of the piston was 194 N for a duration of 10 seconds.

The quality of the lens has been evaluated with a white light interferometer Veeco Wyko NT2000. The images of the surfaces are shown in Figure 8a and 8b. Figure 8c shows the profile of the center of the surface of the lens of the convex side, whereas figure 8d shows the flat side of the lens. The curvature of the lens is too pronounced to measure beyond the central part of the lens using a white light interferometer. The average roughness calculated from the profile is around 0.05 µm for the spherical surface whereas the roughness of the flat surface is 0.088 µm which is acceptable for applications in the mid-infrared. The standard error of this measurement SE=0.02 µm.

The focal length of the lens has been determined using a fiber coupled laser source from Thorlabs at 1550 nm with a power set at 0.05 mW and an infrared camera (Fig.9). The focus in our experiment is at a distance of 15.2 mm from the last surface of the lens. The Back Focal Length, BFL = 11.71 mm which allows to state that the Effective Focal Length, EFL = 11.72 mm. Errors in the horizontal and vertical direction are ±1 µm, and ±0.5 µm respectively.

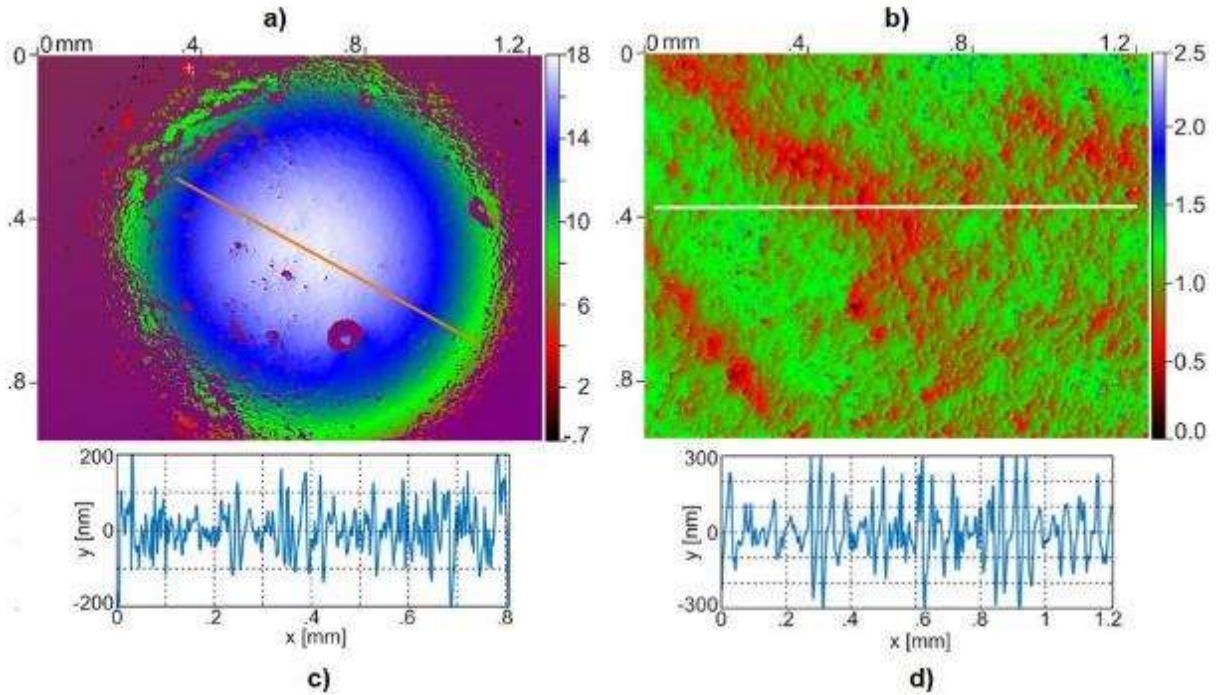

*Figure 8: Surface quality measurements of the developed lens obtained using a white light interferometry method. a) and b) are respectively the scan of the spherical surface and of the flat side of the lens, while c) and d) are the respective surface roughness profiles along the line represented on the scans*

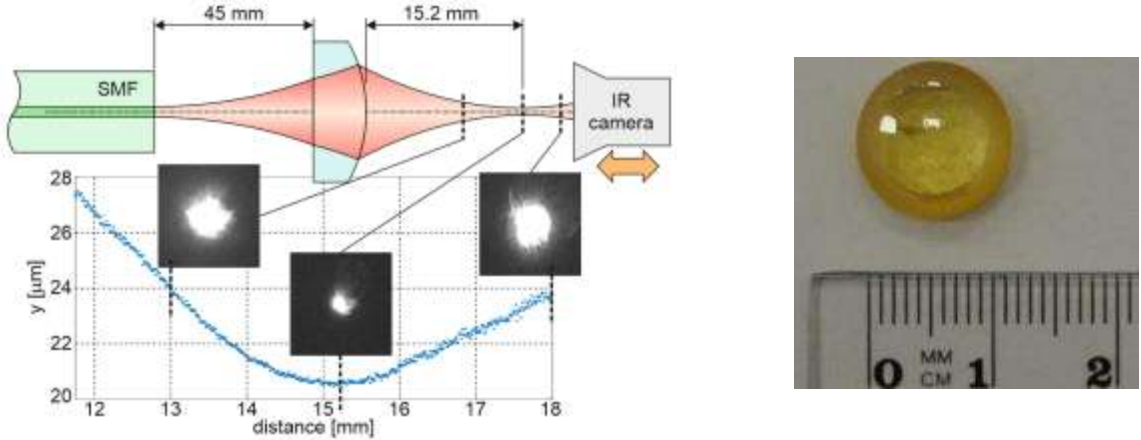

*Figure 9 (left) Experimental setup for the measurement of the effective focal length of the lens, (right) fabricated lens based on SPCG4C glass*

We also verified the possibility to use the selected glass (SPCG4C) for the development of optical fibers by a classic fiber drawing technique considering it's high thermal stability. The preform is fed into a furnace that heat it up to decrease its viscosity, in order to be drawn. Glass fiber is pulled by capstan while the preform is constantly fed into the furnace with constant speed. During the drawing process, fiber is coated by polymer (acrylate) which is UV cured. For this purpose, we first prepared a rod with the dimension 20 x 20 x 200 mm. The rod has then been rounded and polished to make a cylindrical preform (dimension ϕ17.8 x 200 mm). The fabricated optical fiber is composed of a glass core made of SPCG4C while the cladding is made of polyacrylates with a lower refractive index (1.54). The drawing parameters were set as follows:

feeding speed 0.24 mm/min, drawing speed 4 – 6 m/min, temperature 540 – 550ºC. Fibers with core diameter d = 125 µm and 150 µm have been drawn successfully, the overall fiber diameter with the cladding was respectively 225 µm and 250 µm respectively (Fig. 10). We have experimentally proved that this glass could be used as waveguide for future development of optical fibers.

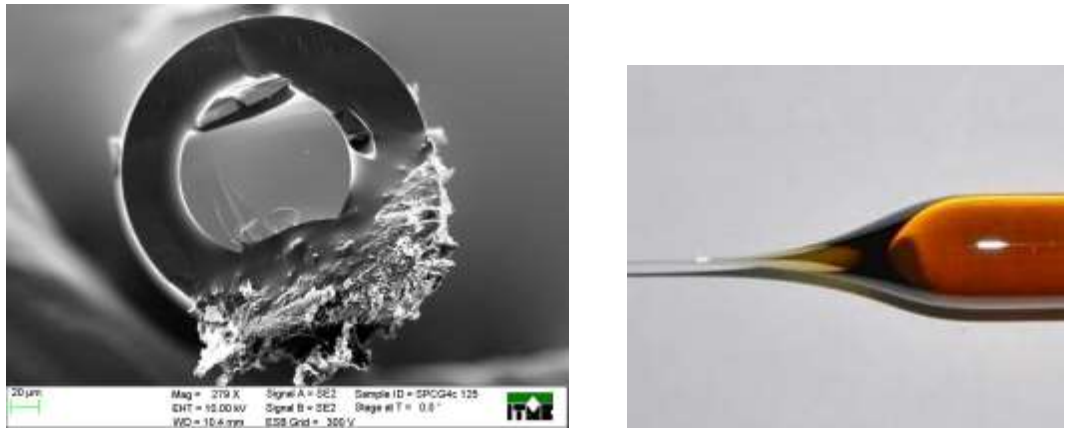

Figure 10 (Left) SEM photo of the fiber made of 125 µm SPCG4C glass core and the polyacrylate cladding. (Right) end of preform of the glass core used to develop the fiber.

**Conclusions**

The $SiO_2$-PbO-CdO-$Ga_2O_3$ oxides system has been studied for the first time as a candidate for crystallization resistant glass for multiple thermal processing. Increasing the gallium oxide content instead of silica tends to increase the thermal stability of the glass against devitrification (as well as increasing the refractive index and the transmission towards mid-infrared). We showed that this system is suitable for the production of optical elements such as lenses and optical fibers due to its high thermal stability. Glass with the composition 28$SiO_2$-45PbO-15CdO-12$Ga_2O_3$ has been selected to develop optical components for its high crystalization resistance and its capacity to be molded. The optimized glass has a transmission up to 5 µm with a refractive index $n_D$ = 1.992. The glass transition temperature $T_g$=460.8 ºC and the softening temperature $T_{sp}$ = 550 ºC, respectively. The thermal stability criterion $\Delta T$=193 K. We have successfully fabricated lens by the hot embossing technique with an average roughness comprised between 40 nm and 90 nm. As a proof of glass feasibility for fiber technology, plastic-clad glass fibers were drawn with diameters of 225 µm and 250 µm and a core diameter of 125 µm and 150 µm.

**Acknowledgement**

This work was supported by European Training Network H2020-MSCA-ITN-2016 Grant No 722380, SUPUVIR: Supercontinuum broadband light sources covering UV to IR applications and by the project TEAM TECH/2016-1/1 operated within the Foundation for Polish Science Team Programme co-financed by the European Regional Development Fund under Smart Growth Operational Programme (SG OP), Priority Axis IV.

Annex 1: Sellmeier coefficients

|  | Sellmeier coefficients | | | | | |
| --- | --- | --- | --- | --- | --- | --- |
|  | B1 | B2 | B3 | C1 | C2 | C3 |
| SPCG1B | 2.2860004 | 0.2676274 | 0.3363165 | 0.0195302 | 0.0862373 | 58.4360 |
| SPCG2B | 2.2934904 | 0.3122492 | 0.0962427 | 0.0144952 | 0.0995960 | 34.5907 |
| SPCG3B | 2.2496469 | 0.4128865 | 1.8550755 | 0.0191284 | 0.0738834 | 207.1074 |
| SPCG4B | 2.2879307 | 0.3843901 | 4.0000000 | 0.0259877 | 0.0576978 | 295.0608 |
| SPCG5B | 2.2832089 | 0.4251855 | 2.0558794 | 0.0190812 | 0.0789427 | 250.0000 |
| SPCG0C | 2.2260472 | 0.4580070 | 0.1151607 | 0.0161229 | 0.0736734 | 21.4515 |
| SPCG1C | 2.2065901 | 0.4374321 | 0.1308677 | 0.0178881 | 0.0749653 | 24.0461 |
| SPCG3C | 2.0058902 | 0.6685237 | 0.2016625 | 0.0144459 | 0.0673399 | 34.2226 |
| SPCG4C | 2.1276564 | 0.5839574 | 1.5320049 | 0.0159745 | 0.0728240 | 214.3810 |
| SPCG5C | 2.5992090 | 0.1272097 | 2.4008329 | 0.0258130 | 0.1132925 | 270.5981 |
| SPCG0D | 2.1585834 | 0.3944738 | 0.2863098 | 0.0167787 | 0.0715992 | 41.2875 |
| SPCG1D | 2.0008162 | 0.5652997 | 2.4000215 | 0.0140470 | 0.0653053 | 297.4703 |
| SPCG2D | 2.3015259 | 0.2754106 | 1.0046092 | 0.0190429 | 0.0844385 | 135.9643 |
| SPCG3D | 2.3975935 | 0.2000000 | 0.2449257 | 0.0213246 | 0.0918612 | 36.9704 |
| SPCG4D | 2.4000000 | 0.2464867 | 0.0657511 | 0.0198281 | 0.0932517 | 17.3433 |
| SPCG5D | 2.0443259 | 0.6266668 | 0.5674441 | 0.0146371 | 0.0691784 | 79.8454 |